\documentstyle[twoside,fleqn,espcrc2]{article}

\newcommand{\be}{\begin{equation}}
\newcommand{\ee}{\end{equation}}
\newcommand{\bea}{\begin{eqnarray}}
\newcommand{\eea}{\end{eqnarray}}

\long\def \omitThis #1 {}
\newcommand{\scsc}{\scriptscriptstyle}
\newcommand{\ts}{\thinspace}
\newcommand{\q}{{\bf q}}
\newcommand{\r}{{\bf r}}
\newcommand{\qh}{{\bf \hat{q}}}
\newcommand{\qs}{{q^\star}}
\newcommand{\alp}{\alpha}
\newcommand{\aq}{\alp_V(q)}
\newcommand{\aqs}{\alp_V(\qs)}
\newcommand{\eps}{\epsilon}
\newcommand{\sig}{\sigma}
\newcommand{\rs}{r_0 \sqrt{\sig}}
\newcommand{\Lam}{\Lambda}
\newcommand{\lam}{\lambda}
\newcommand{\Va}{V_{a^2}(\r)}
\newcommand{\eIR}{e_{{\rm \scsc IR}}}
\newcommand{\plaq}{{\rm plaq}}
\newcommand{\Tr}{{\rm Tr}}
\newcommand{\cO}{{\cal O}}

\newcommand{\ie}{{\it i.e.}}
\newcommand{\eg}{{\it e.g.}}

\title{How to Accurately Extract the Running Coupling of QCD\\
       {}from Lattice Potential Data}

\author{Timothy R. Klassen\address{Newman Laboratory of Nuclear Studies,
        Cornell University, Ithaca, NY 14853, USA}}

\begin{document}

\begin{abstract}
By (a) using an expression for the {\it lattice} potential of QCD in
terms of a {\it continuum} running coupling and (b) globally
parameterizing this coupling to interpolate between 2- (or higher-)
loop QCD in the UV and the flux tube prediction in the IR, we can
perfectly fit lattice data for the potential down to {\it one} lattice
spacing and at the same time extract the running coupling to high
precision. This allows us to quantitatively check the accuracy of
2-loop evolution, compare with the Lepage-Mackenzie estimate of the
coupling extracted {}from the plaquette, and determine the scale $r_0$
ten~times more accurately than previously possible.  For pure SU(3) we
find that the coupling scales on the percent level for $\beta\geq 6$.

\end{abstract}

\maketitle

\section{Introduction}

The string picture predicts that the static potential of QCD behaves
for large distances like
\be\label{VIR}
 V(r) ~=~ \sig r ~-~ \frac{\eIR}{r} ~+~ {\rm const} ~+~ \ldots
\ee
(ignoring string breaking in the presence of dynamical fermions), where
$\sig$ is the string tension and the ``IR charge'' $\eIR = \pi/12$
for the simplest string. In the UV $V(r)$ is Coulomb like,
with a running charge that is known to 2 loops in terms of the
$\Lam$-parameter. The only known general method of obtaining
quantitative information on $V(r)$ is through Monte Carlo (MC)
simulations, which nowadays reach relative accuracies of almost
$10^{-4}, 10^{-3}$ at short, respectively, long distances.  In
contrast, the methods used so far extract $\sig$ {}from these data
with an error of several percent, $\Lam$ with a much larger error.  To
obtain $\sig$ one basically fits to an Ansatz of the form~(\ref{VIR}).
For $\Lam$ one does something different, namely, one defines a running
coupling in the force scheme via $r^2 V'(r) \equiv C_F \alpha_F(r)$
(with $C_F=(N^2-1)/2N$ for SU($N$)), and then fits numerical
derivatives of the potential data to a 2-loop formula for
$\alpha_F(r)$. 2-loop evolution is just becoming good at the shortest
distances presently available, where however lattice artifacts, that
one does not know how to take into account properly, prevent one
{}from seeing it. With these methods one would therefore hardly
benefit {}from more precise data or smaller lattice spacings. Here we
present a method that allows for a unified fit of the whole $r$ range
by incorporating lattice artifacts and the running of the coupling in
a fundamental way.  Details and a complete set of references can be
found in~\cite{TK1}.

\section{The Method}

The general idea is to express the lattice potential in terms of a
continuum running coupling; more precisely there are three ingredients:

\noindent
$\bullet$ Use the V-scheme, where the running coupling is defined via the
continuum static potential
\be\label{alpVdef}
\hat{V}(q) ~\equiv~ -4\pi C_F ~\frac{\aq}{q^2} ~.
\ee
\noindent
$\bullet$ Parameterize $\aq$ to take into account 2-loop QCD in the UV
in terms of $\Lam$, the string prediction~(\ref{VIR}) in the IR in
terms of $\sig$, and to have another parameter for the crossover
{}from UV to IR.

\noindent
$\bullet$ Express the lattice potential as
\be\label{Va}
 \Va = V_0 - 4\pi C_F \int_{-\pi/a}^{\pi/a}\frac{d^3q}{(2\pi)^3}
  {\rm e}^{-i \q \r} ~ \frac{\alpha_V(\hat{q})}{\hat{q}^2}
\ee
where $a \hat{q}_i = 2 \sin(a q_i/2)$ for the usual (unimproved)
gluon action.

The three parameters in $\aq$ and the constant $V_0$ are then fitted
by matching~(\ref{Va}) to the MC data. Though very plausible,
(\ref{Va}) must be considered an Ansatz. The spectacular success of
our fits shows that it incorporates the lattice effects at an
astonishing accuracy. But before presenting these fits we describe the

\section{Parameterization of $\alpha_V(q)$}

The following Ansatz for $\aq$ satisfies the 2-loop $\beta$-function
equation and has no Landau pole (as long as
$c_0\! \geq \! 1, c_1\! \geq \! 1$ and $c \! > \! 0$):
\be\label{alpVansatz}
\frac{1}{\beta_0 \aq} ~=~ \ln\Big[ 1+\frac{q^2}{\Lam^2}
          \ln^b(c_0 +\frac{q^2}{\Lam^2} \lam(q))\Big] ~,
\ee
where $\lambda(q) = \ln^b(c_1 + c \frac{q^2}{\Lam^2})$, and
$b = \beta_1/\beta_0^2$
is a ratio of the first two coefficients of the
$\beta$-function. Of the three dimensionless parameters $c_0, c_1$ and
$c$, the first two are fixed in terms of other parameters by
matching~(\ref{alpVdef}) and~(\ref{alpVansatz}) to~(\ref{VIR}),
\be\label{parfix}
 \ln^b c_0 = \frac{C_F \Lam^2}{2 \beta_0 \sig}, \ts
 \ln^b c_1 = \Big(1-\frac{2\beta_0}{C_F} \eIR \Big)
                              \frac{c_0 \ln^{b+1} c_0 }{2b}
\ee
while $c$ is the crossover parameter to be fitted.

One can show that by iterating the log's in~(\ref{alpVansatz}) in a
suitable way, it is possible, in principle, to incorporate QCD to
{\it any} number of loops, in terms of the (unknown) higher
coefficients of the $\beta$-function and other, non-perturbative
parameters.

\section{Results}
The fitting and error analysis is somewhat involved~\cite{TK1}.
We here only mention this: Since $c$ is quite strongly correlated with
$\Lam$, it is better to use $\aqs$ instead of $\Lam$ as independent
fit parameter. Here $\qs$ is {\it some} UV scale; we chose $\qs =
3.4018/a$ for easy comparison with the Lepage-Mackenzie
estimate~\cite{LM} of $\aqs$ extracted {}from the plaquette $W_{11}$.
We performed fits for various theories, with gauge group SU(3) and
SU(2), without and with fermions (in the latter cases there was no
sign of string breaking).  In all cases we could obtain
$\chi^2/N_{{\rm DF}} \approx 1$ including all points down to $r/a=1$,
which is impossible to achieve with Coulomb $+$ Linear (C$+$L) type
fits.  In fig.~1 we show one of our fits. For our detailed results we
again refer to~\cite{TK1}; in table~1 we compare just two quantities
with previous estimates (we should mention that the $\beta \! = \!
6.8$ data used are preliminary~\cite{BKS}).  One is $\aqs$, the other
the scale $r_0$ \cite{So} defined by $C_F \alpha_F(r_0) \!= \! 1.65$.
[The rhs of this equation is chosen so that $r_0 \approx 0.5~{\rm
fm}$.]  We can use the exact relation between $\aq$ and $\alpha_F(r)$
to calculate $r_0$.  We thereby retain the conceptual advantages of
using a scale like $r_0$ without the errors {}from derivatives of
lattice data.

\begin{figure}[tb]
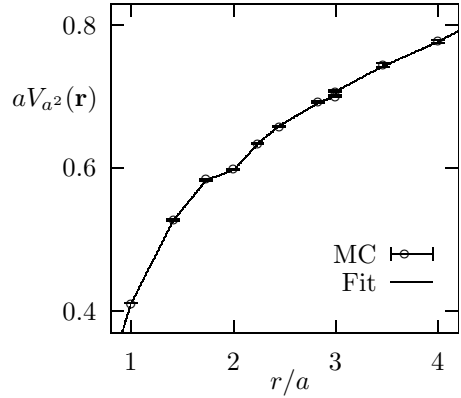

\setlength{\unitlength}{0.240900pt}
\ifx\plotpoint\undefined\newsavebox{\plotpoint}\fi

\caption{Our fit and MC data for $\beta=6$ pure SU(3) in the small
         $r$ region.}
\end{figure}

In fig.~2 we show our running coupling and compare it with estimates
of $\aq$ at various scales extracted by the Lepage-Mackenzie method
{}from various Wilson loops (the rightmost circle corresponding to
$W_{11}$) and Creutz ratios.  We also show the 2-loop approximation to
our $\aq$.

\begin{table*}[hbt]
\setlength{\tabcolsep}{1.245pc}
\newlength{\digitwidth} \settowidth{\digitwidth}{\rm 0}
\catcode`?=\active \def?{\kern\digitwidth}
\caption{Our and previous determinations of $\alpha_V(q^\star)$
         and $r_0$}

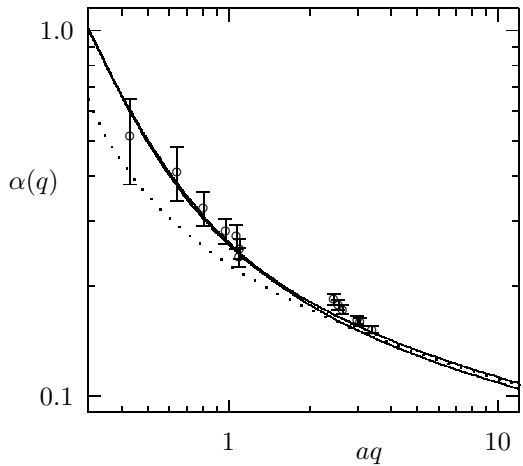
\caption{Fit result for $\aq$ for~$\beta \! = \! 6$ pure SU(3)
         (the solid lines delineating the error),
         its \mbox{2-loop} approximation (dotted), and results {}from
         the method of ref.~\protect\cite{LM} (circles).
         $a^{-1} \! = \! 2.0~{\rm GeV}$. }
\end{figure}

\section{Discussion and Conclusion}

The good agreement between our more precise and the simpler
Lepage-Mackenzie estimate~of $\aqs$ increases the confidence in both
methods.

Our fits show at present hardly any sign of systematic errors.
The only way to check for such errors, then,
is to come up with other parameterizations of $\aq$ that give
similarly good or better
fits. We have done so by taking 3-loop effects
into account. It seems that slight systematic errors exist
only at the ``edges'' of $\aq$, \ie~for the parameters $\aqs$ and
$\sig$. Those of the former are quoted as the second error in table~1.
In the intermediate region $\aq$, and therefore $r_0$,
does not seem to have significant systematic errors.

In contrast to $\sqrt{\sig}/\Lam$, the quantity $\rs$ scales very
well; it equals about $1.14$ for the $n_f=2$ theories of table~1,
$1.17$ for pure SU(2), and 1.17$-$1.19 for pure SU(3).  For the latter
$\rs$ therefore scales at the 1\% level for $\beta\! \geq \! 6.0$. At
the edges of $\aq$ the scaling violations are slightly larger, but we
expect~\cite{BKS} more precise data and analysis to further decrease
them.

As one might suspect {}from fig.~2 and table~1, we are beginning to
gain control of $\aq$ at the fraction of percent level on all momentum
scales (except perhaps in the far IR). For intermediate momenta below
about 5~GeV this is basically already the case; for larger momenta
2-loop evolution is becoming good, so there is hope to soon achieve
this accuracy also in the UV.

The bad scaling properties of quantities involving $\Lam$
are not surprising: We found that the error of
$\Lam$ is quite large --- without leading to a large error in
$\aq$ itself --- because $\Lam$ is strongly correlated with higher
order and non-perturbative contributions at an intermediate range,
appearing here in the form of the fit parameter $c$.

For the future we note that our method can also be used to fit results
{}from improved gluon actions (with the obvious modifications in
eq.~(\ref{Va})), and can be extended to incorporate string breaking.

Finally, it should be clear that our $\aq$ is useful for potential
models of heavy quarks and anywhere one wants to extend perturbation
theory without running into the Landau pole.

I would like to thank Peter Lepage for useful discussions, and the
HEMCGC/SCRI, UKQCD and Wuppertal groups for access to their data. This
work is supported by the NSF.

\end{document}